\documentclass[aps,twocolumn,superscriptaddress,floatfix]{revtex4-2}
\usepackage{graphicx}
\usepackage{amsmath}
\usepackage{amsfonts}
\usepackage{amssymb}
\usepackage{tabularx}

\newcommand{\beq}{\begin{equation}}
\newcommand{\eeq}{\end{equation}}
\newcommand{\beqa}{\begin{eqnarray}}
\newcommand{\eeqa}{\end{eqnarray}}
\newcommand{\eps}{\epsilon}
\newcommand{\rr}{{\bf r}}

\newcommand{\om}{\omega}

\begin{document}
\title{Rotons and their damping in elongated dipolar Bose-Einstein condensates. }
 
\author{S.\,I. Matveenko}
\affiliation{L. D. Landau Institute for Theoretical Physics, Chernogolovka, Moscow region 142432, Russia }
\affiliation{Russian Quantum Center, Skolkovo, Moscow 143025, Russia}

\author{M.\,S. Bahovadinov}
\affiliation{Russian Quantum Center, Skolkovo, Moscow 143025, Russia}
\affiliation{  Physics Department, National Research University Higher School of Economics, Moscow, 101000, Russia}
\author{M.\,A. Baranov}
\affiliation{Center for Quantum Physics, University of Innsbruck, Innsbruck A-6020, Austria}
\affiliation{Institute for Quantum Optics and Quantum Information of the Austrian Academy of Sciences, Innsbruck A-6020, Austria}
 \author{G.\,V. Shlyapnikov}
\affiliation{Russian Quantum Center, Skolkovo, Moscow 143025, Russia}
\affiliation{Moscow Institute of Physics and Technology, Inst. Lane 9, Dolgoprudny, Moscow Region 141701, Russia}
\affiliation{Universit\'e Paris-Saclay, CNRS, LPTMS, 91405 Orsay, France}
\affiliation{Van der Waals-Zeeman Institute, Institute of Physics, University of Amsterdam,Science Park 904, 1098 XH Amsterdam, The Netherlands}

\date{\today}

\begin{abstract}
We discuss  finite temperature damping of rotons in elongated Bose-condensed dipolar gases, which are in the
Thomas-Fermi regime
in the tightly confined directions. The presence of many branches of excitations which
can participate in the damping process, is crucial for the  Landau
damping and results in significant increase of the damping rate. It is found, however, that even
 rotons with energies close to the roton gap may remain fairly stable in systems with the roton gap as small as $1\,nK$.
\end{abstract}
\maketitle

\section{Introduction}

The spectrum of elementary excitations is strongly influenced  by the character of interparticle interactions and is a key
 concept for understanding the behavior of quantum many-body systems. For Bose-condensed systems with a short-range
  interparticle interaction the low-energy part of the  spectrum represents phonons with a linear energy-momentum
  dependence. In some cases, the excitation spectrum has an energy minimum at rather large momenta with roton excitations
   around it, which is  separated by a
  maximum (maxon excitations) from the low-energy phonon part. The roton-maxon excitation was first observed in liquid
   $^4He$, and intensive discussions during  decades arrived at the conclusion that the presence of the roton  is related to the
 tendency to form a crystalline order \cite{he1,he2}.
 The presence  of rotons in the excitation spectrum of dipolar Bose-Einstein condensates  was
  first predicted in Refs. \cite{sant,odel} and is
  considered as a precursor of the formation of a supersolid phase (for a review on the supersolid phase and its
   experimental manifestations see, for example, Ref. \cite{prok}). 
   
During last several years, supersolid phases
    were observed experimentally in systems of ultracold trapped bosonic magnetic atoms (Dy, Er) \cite{ss1,ss2,ss3},
     as well as the presence of the roton excitations and their role in the formation of the supersolid
      state \cite{exp1,exp2,exp3,numericsPRL123}. In systems of magnetic atoms, the formation of the supersolid
       state and the appearance of the roton excitations are attributed to the magnetic dipole-dipole interaction.
        The present theoretical description of the excitations is based on numerical solutions of the
         three-dimensional Bogoliubov - de Gennes equations in a trapped geometry at zero
          temperature \cite{numericsPRL123,num1,exp3}, and is focused on the real part of their dispersion, without
           addressing the question of the excitation damping. The damping, however, strongly affects the system response
            to external perturbations which are used to probe the system properties (see, for example, \cite{Bragg}).
            Therefore, studies of the excitation damping and its temperature dependence have not only theoretical interest,
            but also direct experimental relevance. These studies should also indicate how stable are rotonic excitations, in particular
            for building up roton-induced density correlations in non-equilibrium systems. \cite{ssn2014}. Another issue
            is the spatial roton confinement in trapped Bose-Einstein condensates \cite{santos2013}.

Damping of rotons in quasi-1D dipolar Bose-condensed gases has been  discussed in Refs. \cite{damp1,damp2,damp3}.
In this paper we investigate the damping of  rotons in an elongated  Bose-condensed  polarized dipolar gas, which is
in the Thomas-Fermi (TF) regime in the tightly  confined directions \cite{del1,del2}. In this case, there is a large number of
 branches in the  excitation spectrum, and  many of them can contribute to the Landau damping, which is  the leading damping mechanism
  at finite temperatures  \cite{pit}. This  may significantly increase the damping rate and make rotons unstable.

 The paper is organized as follows. In Section II we present general relations for calculating the condensate wave function, excitation spectrum, and damping rates for rotons. Sections III and IV are dedicated to the approximation of cylindrically isotropic consdensate and contains analytical expressions for excitation energies and the resulting damping rates. In Section V we present the results of direct numerical calculations of these quantities and it is confirmed that the approximation of isotopic condensate gives a qualitatively correct picture. Our concluding remarks are given in Section VI.
\section{General relations}

We consider an elongated Bose-Einstein condensate of polarized dipolar particles (magnetic atoms or polar molecules).
 The motion in the
$z$ direction is free, and  in the $x, y$ directions it is  harmonically confined with frequency $\om$. We consider
 the case where the dipoles are  polarized perpendicularly to the $z$ axis (let  say, are along the $x$ direction). The ground state
  condensate wave function $\Psi_0(\rr)$ obeys  the Gross-Pitaevskii (GP) equation:
\begin{widetext}
\beq
\left[
-\frac{\hbar^2}{2m} \nabla^2 +
\frac{m \om^2 \rho^2}{2} + \int d^3\rr'  V(\rr-\rr') \vert \Psi_0(\rr')\vert^2 \right]\Psi_0(\rr) =\mu \Psi_0(\rr),
\label{GPE}
\eeq
\end{widetext}
where $\rho^2= x^2 + y^2$,  $\mu$ is the chemical potential of the system, and
\beq
V(\rr) = g \delta (\rr) + V_d (\rr),
\eeq
with $g$ being the  coupling constant of the short-range (contact) interaction,
and $V_d(\rr)$ the potential of the dipole-dipole interaction between two atoms. For the dipoles ${\bf d} $ oriented
  in one and the same  direction we have

\beq
V_{d}(\rr) = \frac{d^2 r^2  - 3 (\bf{d} {\rr})^2}{r^5}.
\label{Vd}
\eeq

Representing the field operator of the non-condensed part of the system as
$\Psi'(\rr,t) = \sum_{\nu} [u_{\nu} (\rr) \hat{b}_{\nu} - v_{\nu}^*(\rr) b_{\nu}^{\dagger}  ] $,
where the index $\nu$ labels eigenstates of  the excitations and $b_{\nu}$, $b_{\nu}^{\dagger}$ are their creation and annihilation operators, we have the Bogoliubov-de Gennes equations for the
 functions $u$ and $v$:
\begin{widetext}
\beq
- \frac{\hbar^2}{2 m} \nabla^2 u(\rr)+ \frac{m \om^2 \rho^2}{2} u(\rr) - \mu u(\rr) +[\hat{V}   \Psi^2_0(\rho )] \, u(\rr) + [\hat{V}\Psi_0(\rho) u(\rr)]\, \Psi_0(\rho) - [\hat{V}\Psi_0(\rho) v({\rr}) ]\, \Psi_0(\rho) = E u(\rr),
\label{bdg1}
\eeq
\beq
- \frac{\hbar^2}{2 m} \nabla^2 v(\rr) + \frac{m \om^2 \rho^2}{2}  v(\rr)  - \mu v(\rr) +[\hat{V}  \Psi^2_0 (\rho)]\, v(\rr) + [\hat{V} \Psi_0 (\rho) v(\rr)]\, \Psi_0(\rho) - [\hat{V}  \Psi_0(\rho) u(\rr)] \, \Psi_0(\rho) = -E v(\rr),
\label{bdg2}
\eeq
\end{widetext}
where
\beq
[\hat{V} f(\rr)]  \equiv
\int V(\rr -\rr') f(\rr')   d^3 r' ,
\eeq
for any function $f$, and $E$ is  the excitation energy.

At finite temperatures, the leading damping mechanism for the roton excitation is the Landau damping. In particular,
   a roton  with energy $E_0(q)$ and momentum $q$ interacts with a thermal low-momentum ($|p| \ll q$)
 sound type  excitation with energy $E_j(p)$. Both get annihilated, and an excitation with a higher energy $E_l(q+p)$
is created, where $j,l$ are excitation branch numbers $j, l = 0, 1,2...$.  We  calculate the damping rate for the
 lowest rotonic excitation which has  momentum $k$.     The damping rate is given by the Fermi golden rule
\beq
\frac{1}{\tau}= \sum_{j,l} \frac{1}{\tau_{jl}};
\eeq
\begin{widetext}
\beq
\frac{1}{\tau_{j l}} =  \frac{2 \pi}{\hbar} \int_{-\infty}^{\infty} \frac{d p}{2\pi}
\left[ |\langle q +p, l  | H_{int} |q, \{p, j\}  \rangle |^2 -  |\langle q, \{p, j\}| H_{int} |q + p, l  \rangle|^2 \right]
 \delta (E _{q + p, l} - E_{q, 0} - E_{p, j}).
\label{55}
 \eeq
 The  Hamiltonian $H_{int}$  responsible for the damping represents the interaction between excitations and is given by (see, e. g. \cite{pit})
\beq
H_{int} = \int  d^3\rr \; d^3\rr' \left[  \Psi_0(\rr)  \Psi'^{\dagger} (\rr') V(\rr-\rr') \Psi' (\rr') \Psi_1 (\rr)+\Psi'^{\dagger}(\rr) \Psi'^{\dagger} (\rr') V(\rr-\rr') \Psi' (\rr') \Psi_0 (\rr) \right].
\label{56}
\eeq
\end{widetext}
We use the representation of functions $u, v$ in the non-condensed  operator $\Psi'$  in the form $ u_{\nu} (\rr) = u_{k,j}(\vec{\rho}) e^{i k z}$ and similarly for $v_\nu(r)$, where $k$ is  the wave vector, and $j$ is  the number of the excitation branch.
Then the  Landau damping of rotons  acquires the form
\begin{equation}
\frac{1}{\tau}=\frac{2\pi }{ L\hbar}\sum_{k,n_1,n_2} |A_{k,k+q}^q|^2 (N_k-N_{k+q})\delta(E_{k,n_1}+E_{q,0}-E_{k+q,n_2}),
\label{1tau}
\end{equation}
where     $N_k = 1/(e^{E_k/ T} -1)$    are excitation occupation numbers, and  the matrix element
  $A^q_{k,k+q}$ is the sum of the  integrals:
  \begin{widetext}
\beq
\begin{split}
A^q_{k,k+q}=\int d\rho d\rho^\prime \left[  \Psi_0^*(\rho)  u_{k+q,n_1}^*(\rho^\prime) \tilde{V}(\rho-\rho^\prime,k) u_{k,n_2}(\rho) u_{q,0}(\rho^\prime) +
   \Psi_0^*(\rho) u_{k+q,n_1}^*(\rho^\prime) \tilde{V}(\rho-\rho^\prime,q) u_{q,0}(\rho) u_{k,n_2}(\rho^\prime) \right. \\
+   \Psi_0^*(\rho)  v_{k ,n_1} (\rho^\prime) \tilde{V}(\rho-\rho^\prime,q) u_{q,0}(\rho) v^*_{k+q,n_2}(\rho^\prime) +
  \Psi_0^*(\rho)   v_{q ,0} (\rho^\prime) \tilde{V}(\rho-\rho^\prime,k) u_{k,n_1}(\rho) v^*_{k+q,n_2}(\rho^\prime)\\
+   \Psi_0^*(\rho)  v_{k ,n_1} (\rho^\prime) \tilde{V}(\rho-\rho^\prime,k+q) v^{*}_{k+q,n_2}(\rho) u_{q,0}(\rho^\prime) +
  \Psi_0^*(\rho)   v_{q ,0} (\rho^\prime) \tilde{V}(\rho-\rho^\prime,k+q) v^{*}_{k+q,n_2}(\rho) u_{k,n_1}(\rho^\prime)\\
-   \Psi_0 (\rho^\prime)   u^*_{k+q ,n_1} (\rho )   v_{k,n_2}(\rho^\prime)\tilde{V}(\rho-\rho^\prime,k) u_{q,0}(\rho)
-   \Psi_0 (\rho^\prime)   u^*_{k+q ,n_1} (\rho )   v_{q,0}(\rho^\prime)\tilde{V}(\rho-\rho^\prime,q) u_{k,n_2}(\rho) \\
-  \Psi_0 (\rho^\prime)  v_{k ,n_1} (\rho )   v_{q,0}(\rho^\prime)\tilde{V}(\rho-\rho^\prime,q) v^*_{k+q,n_2}(\rho )
-   \Psi_0 (\rho^\prime)  v_{q ,0} (\rho )   v_{k,n_1}(\rho^\prime)\tilde{V}(\rho-\rho^\prime,q) v^*_{k+q,n_2}(\rho ) \\
-  \left. \Psi_0 (\rho^\prime) v_{k ,n_1} (\rho )   u^{*}_{k+q,n_2}(\rho^\prime) \tilde{V}(\rho-\rho^\prime,k+q)u_{q,0}(\rho )
-  \Psi_0 (\rho^\prime) v_{q ,0} (\rho )   u^{*}_{k+q,n_2}(\rho^\prime) \tilde{V}(\rho-\rho^\prime,k+q)u_{k,n_1}(\rho ) \right].
\end{split}
\label{Aq}
\eeq
\end{widetext}
The Fourier transform of the interaction potential is equal to
\begin{equation}
\tilde{V}(\rho,k)= \int dz V(\rr)e^{ikz}=-\frac{3g\eta}{2\pi}\frac{\partial^2}{\partial^2_x}K_0(k\rho)+g(1-\eta)\delta(\vec{\rho}),
\label{id2}
\end{equation}
with $K_0$ being the modified Bessel function of the second kind, and
\beq
\eta =\frac{4\pi d^2}{3 g}= \frac{g_d}{g}.
\eeq

Generally speaking, the condensate wave function is anistropic in the $x,y$ plane. However, in order to gain insight into the physical picture it is first reasonable to assume that $\Psi^2_0$ is isotropic. This allows us to get analytical expressions for excitations energies and use them for finding the damping rates of rotons. Direct numerical calculations presented in Section V show that the approximation of isotropic condensate gives a qualitatively correct physical picture and reasonable results.

\section{Approximation of isotropic condensate. Excitation spectrum.}
The condensate wave function $\Psi_0$ is $z$-independent and in this section we assume that it depeneds only on $\rho$, i.e it is symmetric in the $x,y$ plane. In the Thomas-Fermi regime the kinetic energy of the condensate is omitted, and one expects that $\Psi_0^2$ has the  shape of  inverted parabola. Then
we obtain
\beq
\int V(\rr -\rr') \Psi_0^2(\rr') d^3 r' \approx g (1+ \eta /2)\Psi_0^2 (\rho).
\label{vx}
\eeq
Equation (\ref{GPE}) then takes the form
\beq
\frac{m \om^2 \rho^2}{2} \psi_0(\rho) + g (1+\eta/2)   \Psi_0^3 (\rho) = \mu \Psi_0(\rho),
\label{TF}
\eeq
and, hence, the condensate wavefunction is given by
 \beq
 \Psi_0^2(\rho) =n_0\left(1-\frac{\rho^2}{R^2}\right) \Theta(R-\rho),
 \label{profile}
 \eeq
 where $\Theta(x)$ is the Heaviside step function, and
 \beq
n_0 = \frac{2}{\pi}\frac{1}{ R^2}n_{1D},
\eeq
with $n_{1D}$ being the  one-dimensional density (the number of particles per unit length in the $z$ direction). The radius of the condensate in the $x, y$ plane is given by
\beq
 R^2 =\frac{2 \mu}{m \om^2}=\frac{2\mu}{\hbar \omega}l_H^2,
 \eeq
 where $l_H=\sqrt{\frac{\hbar}{m\omega}}$ is the harmonic oscillator length, and the relation for $ \mu/\hbar \omega$ is given below in Eq.~(\ref{muHomeg}).
The chemical potential and  density are related to each other as
\beq
 {\mu} = (g+g_d/2)  \frac{2}{\pi R^2}n_{1D}=n_0 g (1+\eta/2),
\eeq
and the validity of the TF regime  requires the  chemical potential (interaction between particles) to be much larger than the level spacing between the trap levels, ${\mu}/{\hbar \om} \gg 1$.
Turning to the functions $ f_{\pm} =u\pm v $ and representing
\beq
f^{\pm} (\rr) = f^{\pm}(\rho) e^{i k z},
\label{0eq}
\eeq
where $k$  is the momentum of the motion along the $z$ axis.  Using the GP equation (\ref{GPE}) we
transform Eqs. (\ref{bdg1}) and (\ref{bdg2}) to
\beq
\frac{\hbar^2}{2m}\left({-\nabla_{{\bf \rho}}^2 + k^2} +\frac{ {\nabla_{\rho}^2}\Psi_0}{\Psi_0}\right) f^+ = E f^-,
\label{1eq}
\eeq
\beq
\frac{\hbar^2}{2m}\left({-\nabla_{{\bf \rho}}^2 + k^2} +\frac{ {\nabla_{\rho}^2}\Psi_0}{\Psi_0}\right) f^- + 2[\hat{V}  \Psi_0 f^-(\rr)]\Psi_0 = E f^+.
\label{2eq}
\eeq

When acting with operator (\ref{id2}) on the function that depends only on $\rho$
it is useful to explicitly differentiate in Eq.(\ref{id2}) and make an average over the azimuthal angle, at least for small and large $k$.
This gives
\beq
\int V(\rr) e^{i k z} dz \approx g(1+\frac{\eta}{2})\delta(\vec{\rho})- \frac{3g\eta}{2\pi}\, \frac{k^2}{2} K_0(k \rho) \equiv A(\rho).
\label{x}
\eeq

In the TF regime we omit  the first and  third  terms  in the round brackets in the l.h.s  of Eq. (\ref{2eq}).
 We then  express $f^-$ through $f^+$ from Eq. (\ref{1eq}) and substitute it into Eq. (\ref{2eq}). This yields
\begin{widetext}
\beq
\frac{\hbar^4 k^2}{4m^2}\left({k^2 -\nabla_{{\bf \rho}}^2 } +\frac{ {\nabla_{\rho}^2}\Psi_0}{\Psi_0}\right) f^+ + 2 \frac{\hbar^2}{2m} \int d^2 r' A(\vert \vec{\rho} -\vec{\rho}' \vert) \Psi_0(\rho')  \left({k^2 -\nabla_{{\bf \rho}'}^2 } +\frac{ {\nabla_{\rho'}^2}\Psi_0}{\Psi_0}\right) f^+(\vec{\rho}')  \Psi_0(\rho) = E^2 f^+(\vec{\rho}).
\label{BdG}
 \eeq
\end{widetext}
In the low momentum limit, $kR \ll 1$, we omit  the first term of Eq. (\ref{BdG}) and angular momentum dependent terms in the
 expression (\ref{x}) for  $A(\rho - \rho')$. Representing $f^+ = W({\rho}) \sqrt{1 - \rho^2/R^2}$, for excitations
  with zero orbital   momentum  (of the  motion around  the $z$ axis) we find
\beq
 (1-\tilde{\rho}^2)(\tilde{k}^2 -\nabla_{\tilde{\rho}}^2) W(\tilde{\rho}) + 2 \tilde{\rho} \frac{d W(\tilde{\rho})}{d \tilde{\rho}}\tilde{k}   = 2 \eps^2  W(\tilde{\rho}),
\label{kll1}
\eeq
where we turned to dimensionless momenta, energy, and coordinates: $\tilde{k}= kR$, $\eps = E/\hbar \om$,  and
$\tilde{\rho} = \rho/R$. In terms of the variable $s = \tilde{\rho}^2$ equation (\ref{kll1}) becomes

\beq
s (1-s) \frac{d^2 W}{ds^2}  +(1-2 s) \frac{d W}{ds} + \left[ \frac{\eps^2}{2}-\frac{\tilde{k}^2}{4} + \frac{s \tilde{k}^2}{4}\right] W=0.
\label{kll2}
\eeq

Omitting the term $\tilde{k}^2 s/4$, Eq. (\ref{kll1}) is nothing else than the hypergeometric equation. This term will be taken into account later in a perturbative approach.
The solution  which is  regular at the origin and finite at $s \to 1$ ($\rho \to  R$)  reads
\beq
W_j = C F(-j, j+1, 1, s ), \quad j = 0, 1, 2, ...,
\label{eps-sound}
\eeq
where  $j$ is  a non-negative integer, and $C$ is the normalization constant.
The related energy spectrum  is given by
$
\eps^2 =  \left[\tilde{k}^2/2 + 2 j (j+1)\right]
$.
From Eq. (\ref{2eq}) we obtain
$\frac{2 \mu}{E} (1-\tilde{\rho}^2)  f^- \approx  f^+$,
and, hence,
  the normalization condition
$ \int d^3 \rr f^{+*} f^- =1$
 gives
\beq
f^+ = \sqrt{\frac{2 \mu}{\hbar \om\, \eps} } \frac{ \sqrt{1-\tilde{\rho}^2}\, W}{\sqrt{\int d^2 \tilde{\rho} W^2}} e^{ikz},
\label{fpq}
\eeq

\beq
f^- = \sqrt{\frac{\hbar \om \, \eps}{2 \mu}} \frac{W}{\sqrt{1-\tilde{\rho}^2} \sqrt{\int d^2 \tilde{\rho} W^2}}e^{ikz}.
\label{fmq}
\eeq
We now take into account the omitted term $\tilde{k}^2 s/4$ perturbatively.  The first order correction to $\eps^2$ is
$\delta \eps^2= -\tilde{k}^2/4$,
for any $j$.
Higher order corrections are proportional to higher powers of $\tilde{k}$ and can be  omitted. Thus, we have
for the spectrum in  the original units
\beq
E_j = \hbar \om \sqrt{(kR)^2 /4+ 2 j (j+1)}.
\eeq
For the lowest branch of the spectrum (j=0) the excitation energy has a linear dependance on $k$:
\beq
E_0= \frac{\hbar \om R}{2} k.
\eeq


In the opposite limit, $k R \gg 1$, we keep the term $(\hbar^2 k_z^2/2m)^2$ in  Eq. (\ref{BdG}).
In this limiting case the main contribution to the integral over $d^2 \rho'$ in equation (\ref{BdG}) comes  from distances
$\vec{\rho'}$ very close to $\vec{\rho}$, and  this equation takes the form (for zero orbital momentum):
\begin{widetext}
\beq
\left( \frac{\hbar^2 k^2}{2m}\right)^2  f^+(\rho)     + 2 g (1-\eta )  \frac{\hbar^2}{2m} \Psi_0  \left({k^2 -\nabla_{{\bf \rho}}^2 } +\frac{ {\nabla_{\rho}^2}\Psi_0(\rho)}{\Psi_0(\rho)}\right) f^+(\rho) \Psi_0(\rho) -6 g \eta  \frac{\hbar^2}{2m} \Psi_0(\rho)
\frac{d^2  f^+ (\rho) \Psi_0}{d\rho^2} = E^2 f^+({\rho}).
\label{rot1}
 \eeq

Representing $f^+ (\rho ) =   (1-\rho^2/R^2) W$, in terms of dimensionless variables $s$, ${\eps}$
equation (\ref{rot1}) reads

\beq
s(1-s) W'' + (1-3s) W' + \left[ \frac{\eps^2 -\eps^2_*(\tilde{k})}{3 \eta}(1+ \eta/2) - \frac{(\eta -1)\tilde{k}^2}{6\eta} s \right] W =0,
\label{rot-exc}
\eeq
\end{widetext}
where
\beq
 \eps_*^2(\tilde{k})=\left(\frac{\hbar  \om}{4 \mu}\right)^2  \tilde{k}^4 -  \frac{1}{2} \frac{\eta -1}{1+ \eta/2} \tilde{k}^2 + \frac{3 \eta}{1+\eta/2}.
\label{eps-rot1}
\eeq
\begin{figure}
   \includegraphics[width=\columnwidth]{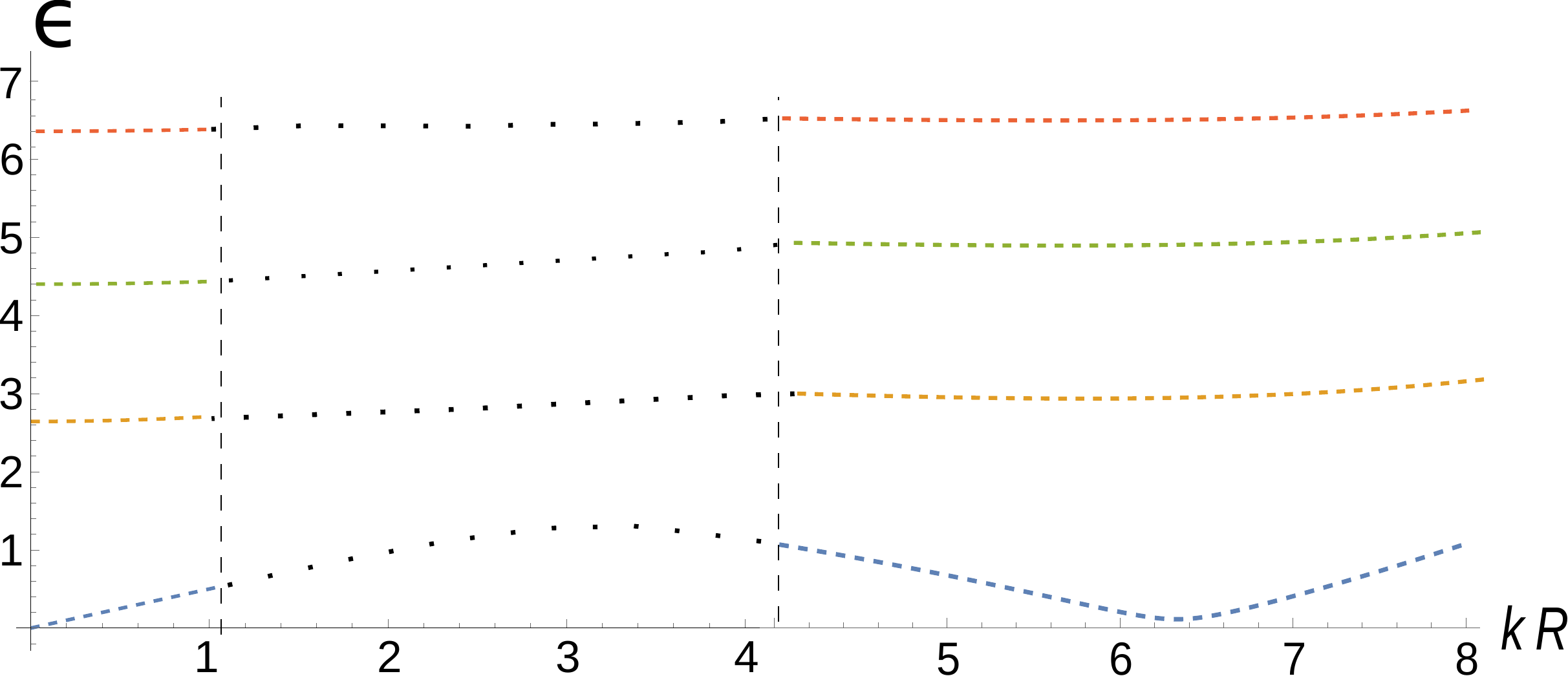}
  \caption{ \label{Fig1} Excitation spectrum  $\eps (\tilde{k})$ as a function of $\tilde{k} = kR$ for $\mu/\hbar \om= 5.9$  ($\beta = 50$), $\eta = 1.8$  ($R/l_H=3.45, \Delta = 1.5nK,  \om =280Hz, \mu =68nK$). Two vertical dashed lines define a region where analytical form of the excitation spectrum does not exist.}
\end{figure}

Omitting the  term $-\frac{(\eta -1)\tilde{k}^2 s}{6 \eta} W$ (which will be taken into account perturbatively later), equation (\ref{rot-exc}) becomes  a hypergeometric equation.
The solution  regular at the origin and finite for $s \to 1$ is
\beq
W_j=  \tilde{C} F\left(-j, j+2, 1, s \right), \,
\label{w-r}
\eeq
where $j$ is  a non-negative integer, and $\tilde{C}$ is  the normalization constant.
 The related energy  spectrum is given by
 $\eps_j^2 = \eps_*^2(q)+ \frac{3 \eta}{1+ \eta/2} j(j+2)$.
 From equations (\ref{1eq}) and (\ref{2eq}) in the limit of $kR \gg 1$ we have

\beq
f^+ \approx \frac{4 \mu \eps}{\hbar \om \tilde{k}^2}f^-,
\label{pm}
\eeq
and, hence,
\beq
f^+ = \sqrt{\frac{4 \mu\eps}{\hbar \om \tilde{k}^2}} \frac{ \sqrt{1-\tilde{\rho}^2}\, W}{\sqrt{\int d^2 \tilde{\rho}\, (1-\tilde{\rho}^2)  W^2}} e^{i k z} ,
\label{fpk}
\eeq
\beq
f^- = \sqrt{\frac{\hbar \om \tilde{k}^2}{4 \mu \eps}} \frac{ \sqrt{1-\tilde{\rho}^2}\, W}{\sqrt{\int d^2 \tilde{\rho}\,  (1-\tilde{\rho}^2) W^2}} e^{i k z}.
\label{fmk}
\eeq

Adding the first order correction to $\eps_j^2$ from the omitted term $-\frac{(\eta -1)\tilde{k}^2 s}{6 \eta} W$, we obtain
\beq
\eps_j^2 = \eps_*^2(\tilde{k})+  \frac{1}{2} \frac{\eta -1}{1+ \eta/2} \tilde{k}^2 x_j + \frac{3 \eta}{1+\eta/2}j (j+2),
\label{eps2}
\eeq
where $x_0 = 1/3$, $x_1= 7/15$, $x_2= 17/35$,  $x_3= 31/63$,... 
 Note that in the considered approximation the roton spectrum exists only at $\eta >1$.
The typical spectrum is shown in Fig.\ref{Fig1} for $\eta = 1.8$. The lowest branch (in blue) has a roton.
 
The character  of the spectrum is determined by two parameters: $\eta$  and $\mu/\hbar \om$. Having in mind direct numerical calculations given below, instead of $\mu/\hbar \om$  we will use the parameter $\beta  =2 n r_*$, where $r_* = md^2/\hbar^2$ is the so called dipolar
length. For isotropic condensate wave function we have  
\beq
\mu/\hbar \om =\sqrt{(2+ \eta)\beta  /3\eta}.
\label{muHomeg}
\eeq

\section{Approximaton of isotropic condensate. Damping of rotons.}
Substituting solutions  (\ref{fpq}), (\ref{fmq}), (\ref{fpk}), (\ref{fmk}) into Eq.~ (\ref{1tau}).
and integrating over  the  momentum $p$ and coordinates,
 we obtain  for the damping rate  of a roton with  momentum $k$     the expression

\beq
\frac{1}{\tau_{jl}} =   \frac{n_0 g^2}{4 \hbar R^2}  \frac{ N_{p, j} - N_{q+p, l}}{\vert E'_{q+p, l}- E'_{p, j}\vert } Z_{j l}, \label{velDif}
  \eeq
\beq
\begin{split}
&Z_{j l} = \left[ G_1(q)  \bar{f}^-_{q,0} f^+_{p,j} \bar{f}^+_{q+p,l} +G_2(q)  \bar{f}^-_{q,0} \bar{f}^-_{p,j} \bar{f}^-_{q+p, l} \right. \\
& + G_3(p)( \bar{f}^-_{p,j}  \bar{f}^+_{q,0} \bar{f}^+_{q+p,l} +
 \bar{f}^-_{p,j}  \bar{f}^-_{q,0} \bar{f}^-_{q+p,l})  \\
&\left. - G_1(q+p) \bar{f}^-_{q+p,l}  \bar{f}^+_{q,0} \bar{f}^+_{p,j} + G_2(q+p)  \bar{f}^-_{q+p,l} \bar{f}^-_{p,j} \bar{f}^-_{q,0}\right]^2, \\
\end{split} \nonumber
 \eeq
where  $E'_{k,l} = d E_{k,l} /dk$,  and
 the momentum $p$ is found from  the energy conservation law:
\beq
E_{q + p, l} = E_{q, 0} + E_{p, j}.
\label{ecl}
\eeq
  The functions $\bar{f}^{\pm}$ are  coefficients  in Eqs. (\ref{fpq}), (\ref{fmq}), (\ref{fpk}), and (\ref{fmk}):
  \[
   \bar{f}^{\pm}_{p, j} = \left[\frac{2 \mu }{\hbar \om \eps_{p, j}}\right]^{\pm \frac{1}{2}},
   \quad  \bar{f}^{\pm}_{q, l}=
   \left[\frac{4 \mu \eps_{q,l}}{ \hbar \om \tilde{q}^2}\right]^{\pm \frac{1}{2}}.
 \]
  For the functions $G_i$ we obtain by the use of Eqs.  (\ref{x}) - (\ref{BdG}) the following expressions:
\beq
\begin{split}
G_1(k)= \frac{(\eps_{k,l}^2 - \left(\frac{\hbar \om}{4 \mu}\right)^2 \tilde{k}^4)(2 + \eta)}{\tilde{k}^2} \\
\times  \frac{ \int_0^1  (1-s) F_1 F_2 ds}{\sqrt{\int F_1^2 ds}\sqrt{\int (1-s) F_2^2  ds}},
\end{split}
\eeq
\beq
\begin{split}
G_2(k)= \frac{(\eps_{k,l}^2 - \left(\frac{\hbar \om}{4 \mu}\right)^2 \tilde{k}^4)(2 + \eta)}{\tilde{k}^2} \\
\times \frac{ \int_0^1 F_1 F_2  ds}{\sqrt{\int (1-s) F_1^2  ds}\sqrt{\int (1-s) F_2^2  ds}},
\end{split}
\eeq
\beq
G_3(p)= (1 + \eta/2)\frac{ \int_0^1 (1-s)  F_1 F_2  ds}{\sqrt{\int F_1^2 ds}\sqrt{\int   (1-s) F_2^2 ds}},
\eeq
where $F_{1,2}$ are the hypergeometric functions:
$ F_1 \equiv F(-j, j+1, 1, s)$, $F_2 \equiv F(-l, l +2, 1, s)$, and we assume in Eqs. (43) - (45)   that $p R \ll 1$, $kR\gg1$. Hence, the quantity $Z_{j l} $ is  a function of the parameters $\eta$ and $\mu/\hbar \om$.

At  temperatures greatly exceeding $\hbar \om$ the occupation numbers are $N_k \approx  T/E_k$, and, consequently, the
 damping rates are linear in $T$:
\beq
\frac{1}{\tau_{jl}} = \frac{k_B T}{\hbar } \frac{m g}{\hbar^2 R}\alpha_{j l}
 Z_{j l},
\eeq
 where
 \beq
 \alpha_{j l} = \left[ \frac{ 1}{\eps_{p,j}} - \frac{1}{\eps_{q+p, l}}\right] \frac{1}{\vert \eps'_{q+p, l}- \eps'_{p, j}\vert}
 \eeq
 and this quantity is responsible for  the increase of the damping  rate in the limit of vanishingly small roton gap
  $\Delta \equiv \eps_0 (q)$. For example, $ \alpha_{00} \propto 1/\Delta$ at $\Delta \to 0$.
For a very small gap the roton excitation is unstable due to strong decay processes.
 Its  energy  becomes of the order of  $\hbar /\tau$ or even smaller.

  We estimate now the damping rate  for some values of
dimensionless parameters $\beta $ and $\eta$ for dysprosium atoms ($r_* =  200\AA $). 

For $\mu/\hbar \om = 7.2$  ($\beta  =70$), $\eta = 1.6$,  $ \om =340Hz$  we have  $  \Delta =0.9 nK, \mu=  101nK,  R = 1.6 \cdot 10^{-4} cm $.   The main decay channel in this case is the one with $j=0$, $l=0$. The damping rate    reaches the value $ \frac{1}{\tau} \sim 10 s^{-1}$ at $T \sim 100nK$.

 A decrease of ${\mu}/{\hbar \om}$ may increase the number of decay channels. In particular, for ${\mu}/{\hbar \om}=5.9$  ($\beta =50$), $\eta = 1.8$,
 $\om =280Hz$, we have  $ \Delta = 1.5nK, \mu =68nK, R = 1.6 \cdot 10^{-4} cm$ and there are two channels: $j=0$, $l=0$; and $j=3$, $l=2$.  The total damping rate  at $T  \sim 100nK$ becomes of the 
  order of $ \frac{1}{\tau} \sim 10^2 s^{-1}$.

For all data   the  damping rate  is much smaller than the roton energy: $\hbar/(\Delta \tau ) \ll 1$. Therefore,  the roton is a well defined excitation.
 
\section{Direct numerical calculations}
\subsection{Condensate wave function and excitation spectrum}
In this section we present our numerical results  for the ground state wave function and excitation spectrum of the condensate.  We numerically solved the GPE equation (\ref{GPE}) for $\Psi_0(x,y)$ using the imaginary-time evolution algorithm in 2D Cartesian grid. The Bogoliubov-de Gennes equations (\ref{bdg1}) - (\ref{bdg2}) for the excitation spectrum are solved using the large-scale Krylov-Schur eigensolver. 

 \begin{figure}
   \includegraphics[width=8cm]{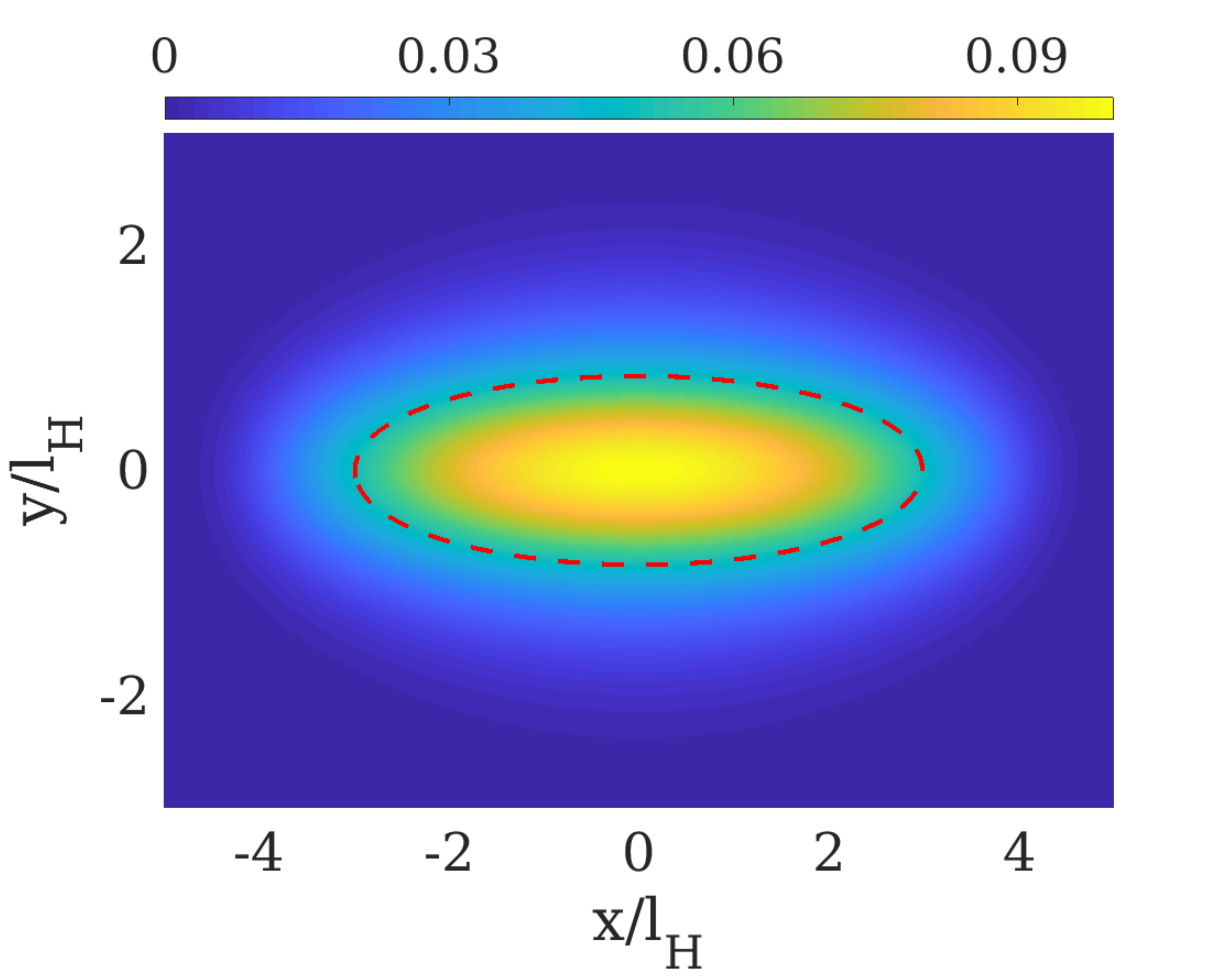}
 
  \caption{ \label{Fig2} The density of the condensate plotted in the $x,y$ plane for $\beta=50$ and $\eta=1.231$. The dashed red line is the countour plot at half-width of the condesate density. The condensate form is anistropic with elongation in the direction of dipoles.    }
\end{figure}
\begin{figure}
   \includegraphics[width=8cm]{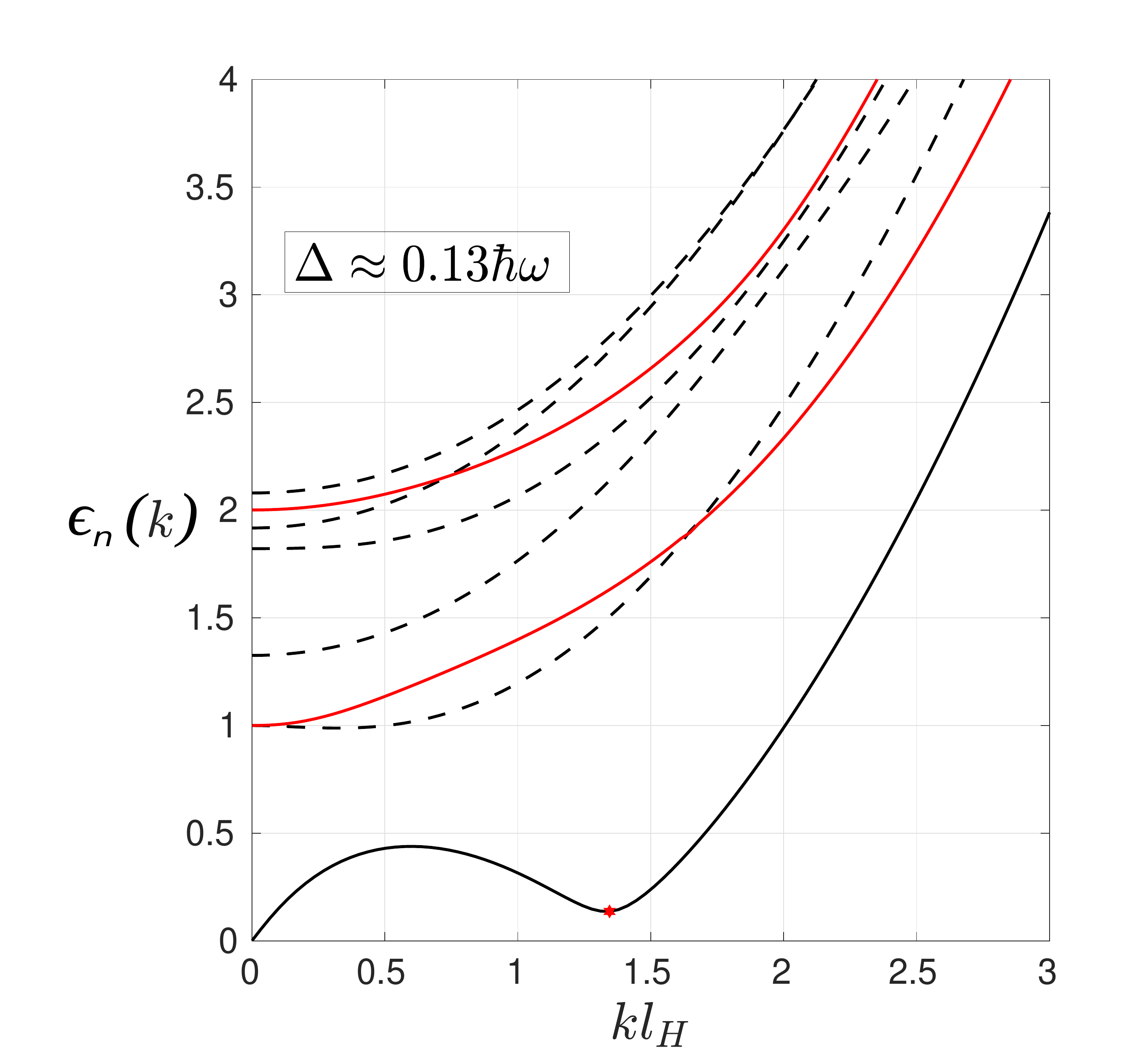}
 
  \caption{ \label{Fig3} Low-lying excitation branches  $\eps_n (k)$  as a function of $ k l_H$ for $\beta=50$ and $\eta=1.231$. The branches are labelled with an index $n$ based on their value at $k=0$ starting from $n=0$. Only the lowest branch $\eps_0 (k)$ has a roton-type excitation with the rotonic gap  $\Delta \approx 0.13 \hbar \omega$. The largest contribution to the roton damping comes from the intraband transitions in the the excitation branches $\eps_2(k)$ and $\eps_6(k)$ (red solid curves).    }
\end{figure}

In Figs.~\ref{Fig2} and \ref{Fig3} we present the condensate density distribution  and the excitation spectrum in the transverse direction for  the dimensionless parameters $\beta=50$ and $\eta=1.231$. In Fig.~\ref{Fig2} the length scale is in units of harmonic oscillator length $l_{H} $.  
 Our numerical results for the condensate wavefunction show its anisotropic form with elongation in the direction of dipoles, as shown in Fig.~2. The red dashed curve in the figure marks the countour plot at half-width of the condesate density.  
 
The excitation spectrum $E_k$ consists of an infinite number of branches. We considered only eight low-lying branches, which is enough for our purposes. As clear from Fig.~\ref{Fig3}, only the lowest band contains a roton type excitation with the rotonic gap $\Delta \approx 0.13 \hbar \omega$  at $k l_H \approx 1.33$ for the dimensionless parameters given above. At a fixed parameter $\beta=50$, increasing/decreasing $\eta$ modifies the excitation spectrum and decreases/increases the rotonic gap $\Delta$. For $\eta \lesssim 1.2$ the rotonic minimum disappears, and there are no rotonic excitations. On the other hand, for $\eta_c \approx 1.2325 $ the rotonic gap goes to zero, and for larger $\eta$ the condensate uniform in the $z$ direction is unstable. For comparison, for a larger value $\beta=104$ the rotonic excitation appears at $\eta \approx 1.12 $ and the rotonic gap vanishes at $\eta_c \approx 1.147$.
%
%
%
 
 \subsection{Damping rate for rotons}
  Using the solutions of the Bogoliubov-de Gennes equations we calculate the damping rate for  the roton excitations with the help of expression (\ref{1tau}).
Due to the broken rotational symmetry, the projection of angular momentum on the $z$-axis is not a conserved quantity and does not serve as a good quantum number. Thus, both intraband ($n_1=n_2$ in Eq.~(\ref{1tau})) and interband ($n_1 \neq n_2$ in Eq.~(\ref{1tau})) transitions can contribute to the damping rate of the rotonic excitation. 

We first present the rates for $\beta=50$ and $\eta=1.231$ and fix the trap frequency $\omega=280 Hz $. Our numerical results show that the main contribution to the damping rate is given by the intraband transition within the third band, i.e the transition with $n_1=n_2=2$ with $1/\tau_{2,2}=75 s^{-1}$ at $T=100nK$. To compare, the damping rate due to the transition in the rotonic branch is $1/\tau_{0,0}=23.9s^{-1}$. Such large contribution of $1/\tau_{2,2}$ results from relatively large matrix elements $A^{q}_{k,k+q}$ and large density of states (vanishingly small velocity difference in Eq.~(\ref{velDif})). The contributions of intraband transitions in higher bands are suppressed due to small matrix elements. As an example, the next largest contribution to the damping rate from the intraband transition is $1/\tau_{6,6} = 4.8 s^{-1}  \ll 1/\tau_{2,2}  $. The total damping rate from all other intraband transitions is smaller than $1/\tau_{6,6}$. 

\begin{figure}[h]
   \includegraphics[width=7.8cm]{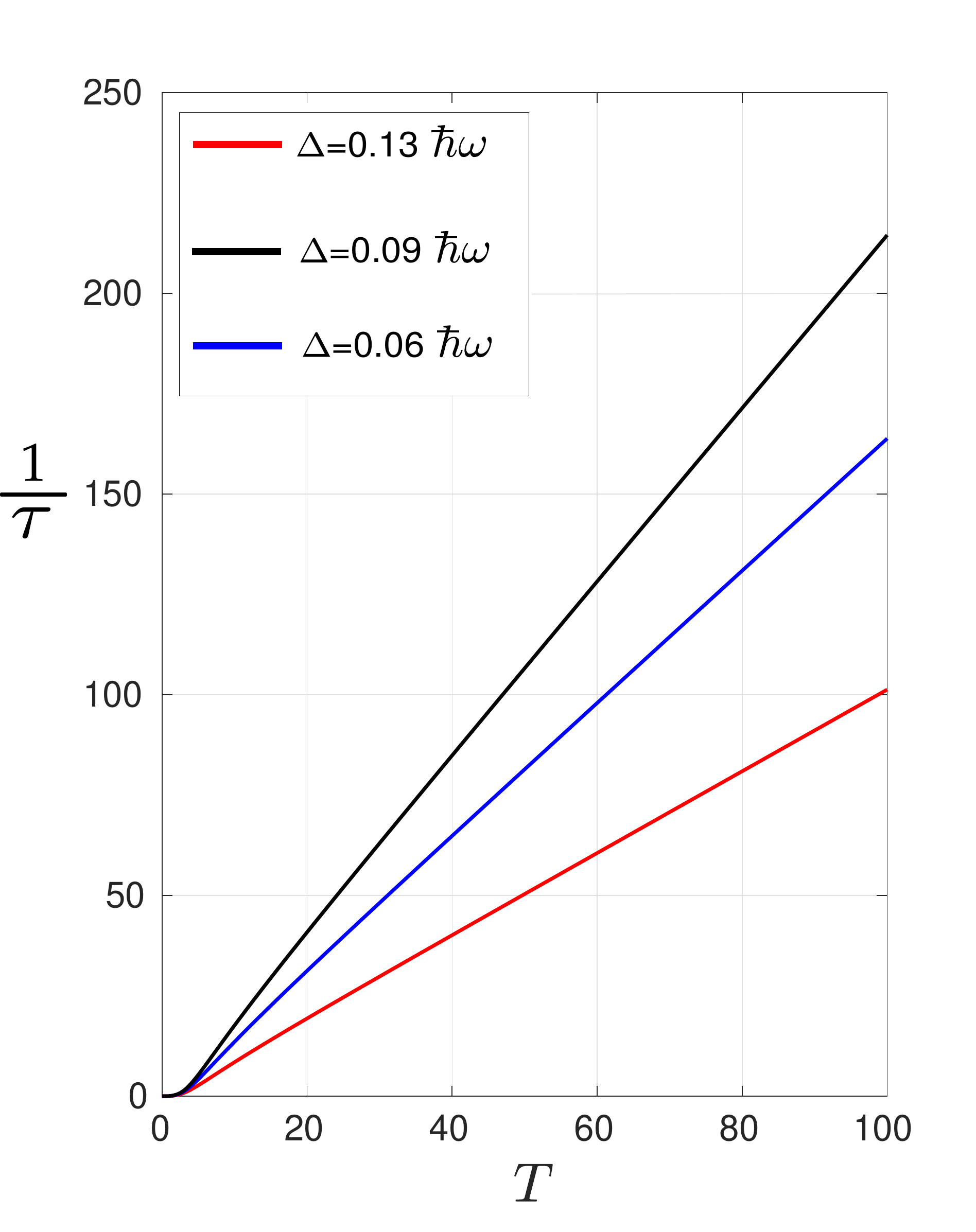}
   \caption{ \label{Fig4} The total damping rates $ 1/\tau $ in units  of $s^{-1}$  versus  temperature $T$ given in $nK$ for $\omega =280 Hz$ and $\beta=50$ in three different regimes with [ $\eta=1.231$, $\Delta=0.13 \hbar \omega$ (red) ], [ $\eta=1.232$, $\Delta=0.09 \hbar \omega$ (blue) ] and [ $\eta=1.2323$, $\Delta=0.06 \hbar \omega$ (black) ]. 
}
\end{figure}

Similarly, our result for the total damping rate due to all possible interband transitions is smaller than $5 s^{-1}$. Although such transitions are not prohibited, practically typical matrix elements are smaller by orders of magnitude.   
 We also considered the damping rates in three other regimes, achieved by varying $\eta$ at $\beta=50$.   We estimated the damping rates at $ \eta=\lbrace 1.225, 1.231, 1.232 , 1.2323 \rbrace $ and presented our final results in Table I. The total damping rate at $T=100nK$ increases with decreasing the rotonic gap from $1/\tau  =20.4 s^{-1}$ at $\Delta=0.28 \hbar \omega$ up to   $1/\tau  =214.5 s^{-1}$ at $\Delta=0.06 \hbar \omega$. Finally, we plot the temperature dependance of the damping rates in Fig.~\ref{Fig4}.
 
            \begin{table}[h]
\caption{\label{tab:rates} Damping rates $\tau_{n,n}^{-1}$ in units of $s^{-1}$ given at $T=100 nK$ at different system parameters $\eta$ with the rotonic gaps $\Delta$. The values $\beta=50$ and $\omega=280 Hz$ are kept fixed.  } 
\begin{ruledtabular}
\begin{tabular}{c|c|c|c|c|c}
    $\eta$ & $\Delta/{\hbar \omega}$  & $\tau_{0,0}^{-1}$ & $\tau_{2,2}^{-1}$ &  $\tau_{6,6}^{-1}$ & $\tau^{-1}$\\[3pt] 
     \hline
       $1.225$&   $0.28$ & $4.9$ & $15.2$ & $0.3$ & $20.4$\\[3pt] 

      \hline
   
      $1.231$ & $0.13$ & $23.9$ & $75$ & $2.4$ & $101.3$\\[3pt] 
 \hline
       $1.232$ & $0.09$ & $29.4$ & $129.6$ & $4.8$ & $163.8$\\[3pt] 
\hline 
     $1.2323$ &  $0.06$ & $30.6$ & $178.2$ & $5.7$ & $214.5$\\[3pt] 
 \end{tabular}
\end{ruledtabular}
\end{table}

\section{Conclusions}
In this paper  we have  calculated the finite temperature damping rate for rotons in an elongated Bose-condensed gas of
 polarized dipolar particles, which is  in the  Thomas-Fermi  regime in the  tightly confined directions. The important feature of this
  case is the presence of a large number of excitation branches which can contribute to the damping process. We found out
   that this leads to a significant increase of the damping rate. Nevertheless, our calculations show that, even in this regime,
    rotons in systems with the roton energy gap of the order of $1\,nK$  are sufficiently long-living and can be observed as
     well-defined peaks in the excitation spectrum and contributions in response functions.

\begin{acknowledgments}
We thank  Francesca Ferlaino    for fruitful discussions. This research was supported  by the Russian Science Foundation
 Grant No. 20-42-05002 and by the joint-project grant from the FWF (Grant No. I4426 RSF/Russia 2019).
  We also acknowledge support of this work by Rosatom.
\end{acknowledgments}


\begin{thebibliography}{99}

\bibitem{he1} E. P. Gross, Classical theory of boson wave fields, Ann.
Phys. (N.Y.){\bf 4}, 57 (1958).
\bibitem{he2}P. Nozi`eres, J. Low Temp. Phys. {\bf 137}, 45 (2004).


\bibitem{odel} D. H. J. O’Dell, S. Giovanazzi, and G. Kurizki, Phys. Rev.
Lett. {\bf 90}, 110402 (2003).

\bibitem{sant} L. Santos, G. V. Shlyapnikov, and M. Lewenstein, Phys.
Rev. Lett. {\bf 90}, 250403 (2003).

\bibitem{prok} M. Boninsegni and N. V. Prokof’ev, Rev. Mod. Phys. {\bf 84},
759 (2012).

\bibitem{ss1} F. Böttcher, J.-N. Schmidt, M.Wenzel, J. Hertkorn, M. Guo,
T. Langen, and T. Pfau, Phys. Rev. X {\bf 9},
011051 (2019).
\bibitem{ss2} L. Tanzi, E. Lucioni, F. Fam`a, J. Catani, A. Fioretti, C.
Gabbanini, R. N. Bisset, L. Santos, and G. Modugno,
 Phys. Rev. Lett. {\bf 122}, 130405 (2019).
\bibitem{ss3} L. Chomaz, D. Petter, P. Ilzhöfer, G. Natale, A. Trautmann, C. Politi, G. Durastante, R. M. W. van Bijnen,
A. Patscheider, M. Sohmen, M. J. Mark, and F. Ferlaino,
 Phys. Rev. X {\bf 9}, 021012 (2019).


\bibitem{exp1} L. Chomaz, R. M.W. van Bijnen, D. Petter, G. Faraoni, S.
Baier, J. H. Becher, M. J. Mark, F. Wächtler, L. Santos, and
F. Ferlaino, Nat. Phys. {\bf 14}, 442 (2018).

\bibitem{exp2} D. Petter, G. Natale, R. M.W. van Bijnen, A. Patscheider, M. J. Mark, L. Chomaz, and F. Ferlaino, Phys. Rev. Lett.
{\bf 122}, 183401 (2019)

\bibitem{exp3} J.-N. Schmidt , J. Hertkorn , M. Guo, F. Böttcher , M. Schmidt, K. S. H. Ng, S. D. Graham, T. Langen,
 M. Zwierlein, and T. Pfau, Roton Excitations in an Oblate Dipolar Quantum Gas, Phys. Rev. Lett. {\bf 126}, 193002, (2021).

\bibitem{numericsPRL123} G. Natale, R. van Bijnen, A. Patscheider, D. Petter, M. J. Mark, L. Chomaz, F. Ferlaino, Phys. Rev. Lett. {\bf 123}, 50402 (2019).

\bibitem{num1} J. Hertkorn, F. Böttcher, M. Guo, J. N. Schmidt, T. Langen, H. P. Büchler, and T. Pfau, Phys. Rev. Lett. {\bf 123}, 193002 (2019).

\bibitem{Bragg}D. Petter,  A. Patscheider, G. Natale, M. J. Mark , M. A. Baranov, R. van Bijnen, S. M. Roccuzzo,
A. Recati, B. Blakie, D. Baillie, L. Chomaz, and F. Ferlaino,
  Phys. Rev. A {\bf 104}, L011302 (2021).

\bibitem{ssn2014} S. S. Natu, L. Companello, and S. Das Sarma, Phys. Rev. A {\bf 90}, 043617 (2014).

\bibitem{santos2013} M. Jona-Lasinio, K. Lakomy, and L. Santos, Phys. Rev. A {\bf 88}, 013619 (2013).

\bibitem{damp1} H. Kurkjian, Z. Ristivojevic,  Phys. Rev. Research {\bf 2}, 033337 (2020).

\bibitem{damp2}  S. S. Natu, R. M. Wilson, Phys. Rev. A {\bf 88} 063638 (2013);  R. M. Wilson S. S. Natu,
 Phys. Rev. A {\bf  93}, 053606 (2016).
\bibitem{damp3}  J. T. Mendonça, H. Terças, A. Gammal,  Phys. Rev. A {\bf 97}, 063610 (2018).

\bibitem{del1} D.H.J. O’Dell, S. Giovanazzi, and C. Eberlein, Phys. Rev.
Lett. {\bf 92}, 250401 (2004).
\bibitem{del2}C. Eberlein, S. Giovanazzi, and D.H.J. O’Dell, Phys.
Rev. A {\bf 71}, 033618 (2005).
\bibitem{pit} L. Pitaevskii and S. Stringari, Bose-Einstein Condensation
and Superfluidity (Oxford University Press, Oxford, 2016),
Vol. 164.

\end{thebibliography}
\end{document}